\documentclass[12pt]{article}
\usepackage{amsmath,amsfonts,amssymb,amscd,amsxtra,amsthm}

\usepackage{graphicx}
\usepackage{epsfig}
\usepackage{bm}
\usepackage{array}

\begin{document}
\title{CKM matrix element $V_{cb}$ in a QCD Potential model}
\author{ $^{1,2}$D K Choudhury and $^{3}$Krishna Kingkar Pathak  \\
$^{1}$Deptt. of Physics, Gauhati University, Guwahati-781014,India\\
$^{2}$ Physics Academy of North-East, Guwahati-781014,India\\
$^{3}$Deptt. of Physics,Arya Vidyapeeth College,Guwahati-781016,India\\
e-mail:kkingkar@gmail.com\\
}
\date{}
\maketitle
\begin{abstract}
  We study the  slope $\rho^{2}$ and curvature $C$ of Isgur Wise function for the heavy-light mesons in general with particular emphasis on $B$ meson in a QCD Potential model. The IW function  is then used to compute the partial decay width and branching ratio for the semileptonic decay of $\left(B^{0}\rightarrow{D,D^{*}l\nu}\right)$. The computed value of the CKM element $|V_{cb}|=0.040$ is found to be in agreement with the available data.   
\end{abstract}

\section{Introduction} 
In a recent communication \cite{kkp5}, we have reported the results of pseudoscalar decay constants of heavy-light mesons in a QCD potential model with linear part of the potential as perturbation. The technique used was the quantum mechanical perturbation theory with plausible relativistic correction.
\\
In this paper, we  extend the  QCD Potential Model with its input parameters towards the semeleptonic decays to study the Isgur-Wise function. Exclusive semileptonic decays of hadrons containing a bottom quark provide a path to measure the Cabibbo-Kobayashi-Maskawa(CKM) matrix elements $V_{cb}$,an important parameter to test the Standard Model. It is well-known in the literature that in case of heavy to heavy transitions like $b\rightarrow c$ decays, all heavy quark bilinear current matrix elements are described in terms of only one form factor, which is called the Isgur-Wise(IW) function in leading order. The IW function, particularly its  slope ($\xi^{\prime}(1)$)at the zero recoil point is important since it allows a model independent way to determine the CKM element $|V_{cb}|$ for the semileptonic decay of $\left(B^{0}\rightarrow{D^{*}l\nu}\right)$ and $\left(B^{0}\rightarrow{Dl\nu}\right)$. This method of obtaining the CKM element within the framework of Heavy Quark Effective Theory (HQET) was proposed by Neubert\cite{neubert1}. He observed that the zero recoil point is suitable for the extraction of CKM element $|V_{cb}|$. The method basically relies on the existence of one universal form factor(Isgur-Wise function) and the fact that the form factor is unaffected from $1/m_{Q}$ $(Q = b, c)$ corrections at zero recoil.\\
 Here, we compute the slope $\rho^{2}$ and curvature $C$ of the Isgur-Wise funcion for $B$ meson and then obtain the CKM element $V_{cb}$.
 In section 2, we discuss the formalism and summarise the results and conclusion in section 3.\\
	     
\section{Formalism}
\subsection{Wave function in the model}
The wavefunction obtained by using the Dalgarno method of perturbation with linear part of the Cornell Potential as perturbation is being discussed in ref.\cite{kkp5,kkp2,NSB2000}. For completeness we write the wavefunction here again.
\begin{equation}
\label{eqn:ch5,1}
\psi_{rel+conf}\left(r\right)=\frac{N^{\prime}}{\sqrt{\pi a_{0}^{3}}} e^{\frac{-r}{a_{0}}}\left( C^{\prime}-\frac{\mu b a_{0} r^{2}}{2}\right)\left(\frac{r}{a_{0}}\right)^{-\epsilon}
\end{equation}
\begin{equation}
\label{eqn:ch5,2}
N^{\prime}=\frac{2^{\frac{1}{2}}}{\sqrt{\left(2^{2\epsilon} \Gamma\left(3-2\epsilon\right) C^{\prime 2}-\frac{1}{4}\mu b a_{0}^{3}\Gamma\left(5-2\epsilon\right)C^{\prime}+\frac{1}{64}\mu^{2} b^{2} a_{0}^{6}\Gamma\left(7-2\epsilon\right)\right)}}
\end{equation}
\begin{equation}
\label{eqn:ch5,3}
C^{\prime}=1+cA_{0}\sqrt{\pi a_{0}^{3}}
\end{equation}
\begin{equation}
\label{eqn:ch5,4}
\mu=\frac{m_{i}m_{j}}{m_{i}+m_{j}}
\end{equation}
\begin{equation}
\label{eqn:ch5,5}
a_{0}=\left(\frac{4}{3}\mu \alpha_{s}\right)^{-1}
\end{equation}
\begin{equation}
\label{eqn:ch5,6}
\epsilon=1-\sqrt{1-\left(\frac{4}{3}\alpha_{s}\right)^{2}}.
\end{equation}
The QCD potential is taken as
\begin{equation}
\label{eqn:ch5,7}
V\left(r\right)=-\frac{4}{3r}\alpha_{s}+br+c
\end{equation}
Here $A_{0}$ is the undetermined factor appearing in the series solution of the Schr\"odinger equation.However with $A_{0}$=0 the effect of c vanishes entirely from the solution.  \\
The corresponding normalised wavefunction in momentum space is\cite{kkp5}:
\begin{equation}
\label{eqn:ch5,8}
\psi_{rel+conf}\left(p\right)=\frac{N^{\prime}\sqrt{2}\left(2-\epsilon\right)\Gamma\left(2-\epsilon\right)}{\pi\left(1+ a_{0}^{2}p^{2}\right)^{\frac{3-\epsilon}{2}}}\left[C^{\prime}-\frac{\left(4-\epsilon\right)\left(3-\epsilon\right)\mu b a_{0}^{3}}{2\left(1+a_{0}^{2}p^{2}\right)}\right].
\end{equation}

\subsection{The Isgur-Wise function and Semileptonic decay}
In case of weak decay like the semileptonic $B\rightarrow Dl\nu$ decay, the amplitude is a product of two matrix elements, the hadronic weak current$H_{\mu}$ and the leptonic weak current$L_{\mu}$ with its vector $V^{\mu}$ and axial vector $A^{\mu}$ i.e
\begin{equation}
A=\frac{G_{F}}{\sqrt{2}}V_{cb}\langle l^{-}(k_{1}),\overline{\nu_{l}}(k_{2})|L_{\mu}|0\rangle \times \langle D(P^{\prime})|H^{\mu}|\overline{B}(p)\rangle
\end{equation}
where $H_{\mu}=\overline{c}\gamma^{\mu}(1-\gamma_{5})b$ and $L_{\mu}=\overline{l}\gamma^{\mu}(1-\gamma_{5})\nu_{l}$.

Since both the $B$ and $D$ mesons are pseudoscalar $(J^{p}=0^{-})$, so from the consideration of lorentz covariance and parity one has $ \langle D(P^{\prime})|A^{\mu}|\overline{B}(p)\rangle =0$ and there remains the vector part
 
 \begin{equation}\label{eq.p}
\langle D(p^{\prime})|V^\mu|B(p)\rangle=f_+(q^2)(p+p^{\prime})_\mu+f_-(q^2)(p-p^{\prime})_\mu. 
\end{equation}
 where $f_+(q^2)  $ and$ f_-(q^2) $ are two form factors.\\
 Similarly for $B\rightarrow D^{*}l\nu$, we get another four independent form factors
 \begin{eqnarray}\label{eq.v}
\langle D^{*}(p^{\prime},\epsilon)|\bar u\gamma^\mu b|B(p)\rangle&=&2i\epsilon^{\mu\nu\alpha\beta}\frac{\epsilon_\nu p^{\prime}_\alpha p_\beta}{M_{B}+M_{D^{*}}}V(q^2)\\
\langle D^{*}(p^{\prime},\epsilon)|\bar u\gamma^\mu\gamma_5b|B(p)\rangle&=&(M_{B}+M_{D^{*}})\left[\epsilon^\mu-\frac{\epsilon\cdot qq^\mu}{q^2}\right]A_1(q^2)\nonumber\\
&&-\epsilon\cdot q\left[\frac{(p+p^{\prime})^\mu}{M_{B}+M_{D^{*}}}-\frac{(M_{B}-M_{D^{*}})q^\mu}{q^2}\right]A_2(q^2)\nonumber\\
&&2M_{D^{*}}\frac{\epsilon\cdot q q^\mu}{q^2}A_0(q^2)
\end{eqnarray}
 Unlike the case of electromagnetic interaction here the normalisation of weak form factors are in general unknown.\\
 However, in the limit of infinitely heavy quark masses $M_{B}\rightarrow\infty$, a new heavy flavour symmetry appears in the effective Lagrangian of the standard model which provides the model independent normalisation of the weak form factors in the framework of HQET. \\ 
 In the heavy-quark limit the masses of the heavy quarks and consequently, the masses of the heavy hadrons are taken to be infinite. This leads to an additional symmetry which is known as the heavy flavour symmetry. With the hadron masses their momenta also go to infinity but the hadron four velocities remain finite in this symmetry. One is then interested in the dependence of form factors on the (finite) velocity product $v.v^{\prime}$ . Moreover, the heavy quark symmetry is an approximate symmetry and correction arises since the quark masses are not infinite. This correction  may be studied systematically in the framework of HQET. The leading symmetry-breaking corrections are from the terms of the order of $\frac{1}{m_{Q^{*}}}$ .
Although the relative corrections can be calculated using perturbative QCD, the $m_{Q^{*}}$ corrections produce new uncalculable functions. For $B\rightarrow D^{*}l\nu$ decay there are found to be four such uncalculable functions. Hence the predictive power of the theory is reduced. Isgur, Wise, Georgi and others\cite{neubert1,Neubert,Isgur} showed that in case of weak semileptonic decays of $B\rightarrow Dl\nu$ or $B\rightarrow D^{*}l\nu$, all the form factors are expressible in terms of a single universal function of velocity transfer, which is normalized to unity at zero recoil. This universal function is known as the Isgur-Wise function, which measures the overlap of the wave functions of the light degrees of freedom in the initial and final mesons moving with velocities $v$ and $v^{\prime}$ respectively.\\
\subsection{Isgur-Wise function in the model}

In the heavy quark limit close to $y=1$, the Isgur-Wise function is written as  :
\begin{eqnarray}
\label{eqn:ch5,IW}
\xi\left(v_{\mu}.v^{\prime}_{\mu}\right)\nonumber&=&\xi\left(y\right)\\&=&1-\rho^{2}\left(y-1\right)+ C\left(y-1\right)^{2}+...
\end{eqnarray}
where 
\begin{equation}
y= v_{\mu}.v^{\prime}_{\mu}
\end{equation}
and $v_{\mu}$ and $v^{\prime}_{\mu}$  being the four velocity of the heavy meson before and after the decay.The quantity $\rho^{2}$  is the slope of I-W function at $y=1$ and known as charge radius :\\

\begin{equation}
\rho^{2}= \left. \frac{\partial \xi}{\partial y}\right.|_{y=1}
\end{equation}
The second order derivative is the curvature of the I-W function known as convexity parameter :\\

\begin{equation}
C=\left .\frac{1}{2}\right. \left(\frac{\partial^2 \xi}{\partial y^{2}}\right)|_{y=1}
\end{equation}
For the heavy-light flavor mesons the I-W function can also be written as \cite{NSB2000,close} :\\

\begin{equation}
\label{eqn:ch5,IW2}
\xi\left(y\right)=\int_{0}^{+\infty} 4\pi r^{2}\left|\psi\left(r\right)\right|^{2}\cos pr dr
\end{equation}
where\\

\begin{equation}
p^{2}=2\mu^{2}\left(y-1\right)
\end{equation}
Upon integration of eqn.\ref{eqn:ch5,IW2} one obtains\cite{NSB2000,NSB2009}
\begin{eqnarray}
\label{eqn:ch5,IW1}
\xi\left( y \right)= 1-\frac{N^{\prime2}a_{0}^{2}\mu^{2}\left(y-1 \right)}{2}\left[\frac{4\Gamma\left(5-2\epsilon\right)}{2^{4-2\epsilon}}-\frac{4\mu b a_{0}^{3}\Gamma\left(7-2\epsilon\right)}{2^{6-2\epsilon}}+\frac{\mu^{2} b^{2} a_{0}^{6}\Gamma\left(9-2\epsilon\right)}{2^{8-2\epsilon}}\right]  \nonumber\\ + \frac{4N^{\prime2}a_{0}^{4}\mu^{4}\left(y-1 \right)^{2}}{2^{5-2\epsilon}} 
\left[ \Gamma\left(4-2\epsilon\right)\left\{\frac{5-2\epsilon}{8} + \frac{3-2\epsilon}{3}+ \frac{\left(3-2\epsilon \right)^{2}}{4} + \frac{\left(3-2\epsilon\right)^{3}}{24} \right\}\right. \nonumber\\ \left.- \frac{\mu b a_{0}^{3}\Gamma\left(6-2\epsilon\right)}{4}\left\{\frac{7-2\epsilon}{8} + \frac{5-2\epsilon}{3}+ \frac{\left(5-2\epsilon \right)^{2}}{4} + \frac{\left(5-2\epsilon\right)^{3}}{24} \right\}\right. \nonumber \\ \left. +\frac{\mu^{2} b^{2} a_{0}^{6}\Gamma\left(8-2\epsilon\right)}{64}\left\{\frac{9-2\epsilon}{8} + \frac{7-2\epsilon}{3}+ \frac{\left(7-2\epsilon \right)^{2}}{4} + \frac{\left(7-2\epsilon\right)^{3}}{24} \right\}\right]. 
\end{eqnarray}
The normalized wave function naturally follows the zero recoil condition of the IW function $\xi(1)=1$ in the model. \\

\subsection{The strong coupling constant $\alpha_{s}$ in the Model}
In the perturbation procedure, the convergence point of view demands from the eqn.\ref{eqn:ch5,8} that
 \begin{equation}
 \label{eqn:ch5,9}
 \frac{(4-\epsilon)(3-\epsilon)\mu b a^{3}_{0}}{2(1+a^{2}_{0}p^{2})}<<C^{\prime}.  
\end{equation}
This condition provides a lower limit of $\alpha_{s}\geq 0.38$ in the model to incorporate a small value of $p_{0}^{2}$($p_{0}^{2}\leq \Lambda^{2}$). Again the reality condition of $\epsilon$ provides an upper limit of $\alpha_{s} \leq \frac{3}{4}$. Hence the condition of equation.(\ref{eqn:ch5,9}) allows a range of $0.38\leq \alpha_{s}\leq 0.75$ in the model to treat the linear part of the Potential as perturbation with its model parameters.\\
 Instead of using a free strong coupling constant $\alpha_{s}$ within this range, we therefore consider
the strong running coupling constant appeared in the potential $V(r)$, is related to the quark mass parameter as\cite{kkp5,kkp2,badalianalpha,Faustovalpha} 
\begin{equation}
\label{eqn:Faustovalpha}
\alpha_{s}\left(\mu^{2}\right)=\frac{4\pi}{\left(11-\frac{2n_{f}}{3}\right)ln\left(\frac{\mu^{2}+M^{2}_{B}}{\Lambda^{2}}\right)}
\end{equation}
where, $n_{f}$ is the number of flavours, $\mu$ is the renormalisation scale related to the constituent quark masses as $\mu=2\frac{m_{i}m_{j}}{m_{i}+m_{j}}$ and $M_{B}$ is the background mass. The background mass is not an arbitrary parameter and can be calculated in the frame work of lattice QCD. The apperance of mass $M_{B}$ in eqn.\ref{eqn:Faustovalpha} is similar to the case of QED where $\alpha$ has the mass of $e^{+}e^{-}$ pair under logarithm \cite{badalianalpha}. Here we relate the background mass to the confinement term of the potential as $M_{B}=2.24 \times b^{1/2}=0.95 GeV$ \cite{Faustovalpha,kkp5}.\\
In studying leptonic decay \cite{kkp5}, the value of $\Lambda_{QCD}=200 MeV$ was fixed in the eqn.\ref{eqn:Faustovalpha} to obtain the value of running background coupling $\alpha_{s}$. In the case of leptonic decay of charged mesons, the quark and antiquark annihilate to produce a virtual $W^{\pm}$ boson such that $q^{2}=M^{2}$ and hence we get only one form factor, the decay constant  $f_{p}$ which absorb all the strong interaction effects. In the case of semileptonic decay, however the case is different since $q^{2}$ is different for event to event and hence more than one form factor appears. This decrease of $q^{2}$ in semileptonic decay leads us to consider a larger value of $\Lambda_{QCD}$ in compared to that of the leptonic decay which effectively increases the strong coupling constant $\alpha_{s}(\mu)$. The physically plaussible range of effective $\Lambda_{QCD}$ should in principle be deduced from the allowed bands of the slope and curvature of the I-W function. Considering the theoretical bounds on slope  $3/4\le\rho^{2}<1.51 $\cite{Jugeau,younac} and curvature $C\ge\frac{5\rho^{2}}{4}$\cite{younac}, we draw a curve for the variation of I-W function for $B$ meson and is shown in fig.1. The allowed range of $\rho^{2}$ provides a range of $\Lambda_{QCD}$ in the model as $382 MeV\le\Lambda_{QCD}\le430 MeV$.\\
\begin{figure}[htb]
\begin{center}
\includegraphics[scale=0.8]{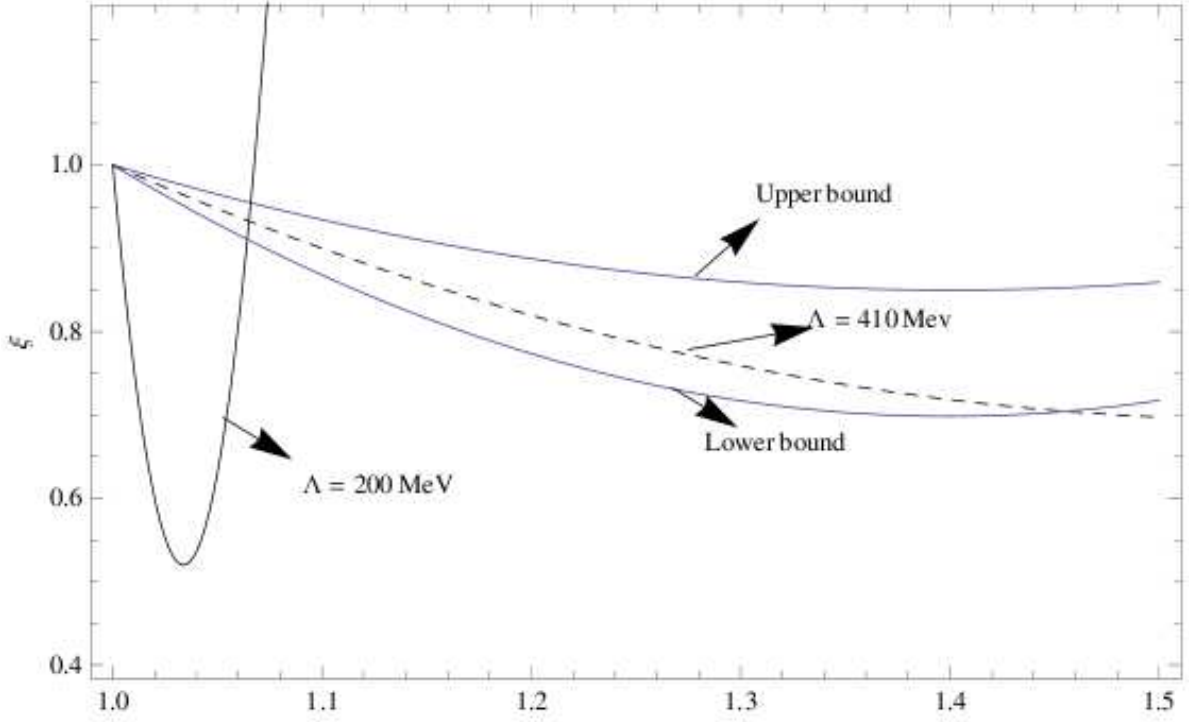}\label{fig:ch5,iw}
\caption {Variation of I-W function with Y for different scales of $\Lambda$.}
\end{center}
\end{figure}
 In ref.\cite{badalianalpha}, it was analysed in studying the freezing of QCD coupling effects that for running background coupling in $V$ scheme, one can choose $\Lambda_{V}(n_{f}=3)=410 MeV$(eqn.30 of ref.\cite{badalianalpha}) and this value of $\Lambda_{V}$ does not contradict with those which are commonly used in $\overline{MS}$ renormalisation scheme and give rise to $\alpha_{s}(M_{Z})=0.118\pm0.001$. Hence we justify ourself to choose $\Lambda_{QCD}=0.410GeV$ in this work.
 The input parameters used in the numerical calculation are the same as is used in our previous works\cite{kkp5} $n_{f}=3$, $m_{d}=0.336 GeV$, $m_{c}=1.55 GeV$, $m_{b}=4.97 GeV$ with $b=0.183$ $GeV^{2}$ and $cA_{0}$=1$GeV^{2/3}$ with $c=-0.4 GeV$\\

\section{Results}

\subsection{Determination of CKM element $V_{cb}$}
Using eqn.\ref{eqn:ch5,IW} and eqn.\ref{eqn:ch5,IW1}, we compute the slope and curvature of the I-W function for $B$  meson. 
It is also added that the result with $\Lambda_{QCD}=410$ MeV is found to be improved than our previous work with $V$ Scheme\cite{NSB2009}.
\begin{table}[h]
\begin{center}
\caption{Slope and curvature of I-W function for $B$ and $D$ mesons.}\label{tab:ch5,1}
\vspace{0.2cm}
\begin{tabular}{|c|c|c|}
\hline Slope and Curvature  & With $\Lambda_{QCD}=200 MeV$ & With $\Lambda_{QCD}=410 MeV$ \\ \hline
 $\rho^{2}_{B}$ &  28.42 &  1.11\\ 
$C_{B}$ & 420.68 & 0.99 \\ 
\hline 
\end{tabular} 
\end{center}
\end{table}

 From the table.\ref{tab:ch5,1}, we see that with $\Lambda_{QCD}=200$ MeV, the results overshoots all other theoretical upper bounds $3/4\le\rho^{2}<1.51 $\cite{Jugeau,younac} and hence for determination of $V_{cb}$ we consider the value of $\rho^{2}_{B}$ and $C_{B}$ with $\Lambda_{QCD}=410 MeV$. 
 
\begin{table}
\begin{center}
\caption{ Comparison of slope and curvature of $B$ mesons with other works.}\label{tab:ch5,2}
\vspace{0.2cm}

\begin{tabular}{|c|c|c|}\hline
  work  & $\rho^{2}_{B}$  & $C_{B}$ \\\hline   
Present & 0.993          & 1.114   \\
Faustov etal;\cite{Faustovnewanalysis}& 1.04&1.36\\
Lattice QCD\cite{Bowler}&$0.83^{+15+24}_{-11-22}$&..\\
ALEPH\cite{ALEPH}&$0.92\pm0.98\pm0.36$&..\\
Belle\cite{Belle}&$1.12\pm0.22\pm0.14$&..\\
Le Youanc et al \cite{younac} &$\ge 0.75$&$\ge 0.47$\\ 
QCD Sum Rule \cite{BDAI}&0.65&0.47\\
Relativistic Three Quark Model \cite{IVANOV}&1.35&1.75\\
Neubert \cite{NEUBERT}&0.82$\pm$0.09&..\\\hline
\end{tabular}

\end{center}
\end{table}

The differential semileptonic decay  rate $B\to D\,l\,\bar{\nu}$ for
the massless leptons is given by\cite{neubert1,Qkim}
\begin{equation}
\frac{d\Gamma}{d y}=\frac{G_F^2}{48\pi^3}\,|V_{cb}|^2\,M_D^3\,
(y^2-1)^{3/2}(M_B+M_D)^2\ \xi^{2}\left(  y \right),                                       
\end{equation}
where $y$ lies in the range $1\le y\le \frac{M^{2}_{B}+M_{D}^{2}-m^{2}_{l}}{2M_{B}M_{D}}$.\\
 The differential semileptonic decay rate $B\to
D^*\,l\,\bar{\nu}$ is defined by 
\begin{eqnarray}
\frac{d\Gamma}{d y}&=&\frac{G_F^2}{48\pi^3}\,|V_{cb}|^2\,(M_B-M_{D^*})^2\,M_{D^*}^3\,
\sqrt{(y^2-1)}\,(y+1)^2\cr
&&\times\bigg[
1+\frac{4y}{y+1}\ \frac{q^{2}\left(y\right)}{(M_{B}-M_{D^{*}})^2}
\bigg]\ \xi^{2}\left(  y \right),                     
\end{eqnarray}
where $q^{2}(y)=M^{2}_{B}+M^{2}_{D^{*}}-2 y M_{D^{*}}M_{B}$.
By integrating the expressions for the differential decay rates
\begin{equation}
\Gamma=\int\frac{G_F^2}{48\pi^3}\,|V_{cb}|^2\,M_D^3\,
(y^2-1)^{3/2}(M_B+M_D)^2\ \xi^{2}\left(  y \right)d y\\
\end{equation}
and
\begin{eqnarray}
\Gamma&=&\int\frac{G_F^2}{48\pi^3}\,|V_{cb}|^2\,(M_B-M_{D^*})^2\,M_{D^*}^3\,
\sqrt{(y^2-1)}\,(y+1)^2\cr
&&\times\bigg[
1+\frac{4y}{y+1}\ \frac{q^{2}\left(y\right)}{(M_{B}-M_{D^{*}})^2}
\bigg]\ \xi^{2}\left(  y \right)dy,                     
\end{eqnarray}
 we get the predictions for the total decay rates in our model as
\begin{eqnarray}
\Gamma(B\to D l\nu)&= &6.82|V_{cb}|^2\ {\rm ps}^{-1},\\
\Gamma(B\to D^* l\nu)&=&28.40|V_{cb}|^2\ {\rm ps}^{-1}.
\end{eqnarray}

Taking mean values of lifetimes from PDG2012:
$\tau_{B^0}=1.519\times 10^{-12}$~s and $\tau_{B^+}=1.641\times
10^{-12}$~s, and also using the experimental masses for the eq.24 and eq.25, we compute the semileptonic branching ratio by using the relation $BR=\Gamma\times \tau$ and find
 \begin{eqnarray}
  BR(B^0\to D^+ l^-\nu)&=&10.401|V_{cb}|^2,\cr
BR(B^+\to D^0 l^+\nu)&=&11.10|V_{cb}|^2,\cr
 BR(B^0\to D^{*+} l^-\nu)&=&43.31|V_{cb}|^2,\cr
BR(B^+\to D^{*0} l^+\nu)&=&46.78|V_{cb}|^2.
\end{eqnarray}
The comparison of these theoretical results with the experimental branching ratios from the lattest PDG \cite{pdg2012}
 gives us following values of the CKM matrix element $|V_{cb}|$: 
 \begin{eqnarray}
   BR(B^0\to D^+ l^-\nu)^{\rm exp}=0.0218&\qquad& |V_{cb}|=0.046,\cr
BR(B^+\to D^0 l^+\nu)^{\rm exp}=0.0229&\qquad& |V_{cb}|=0.045,\cr
 BR(B^0\to D^{*+} l^-\nu)^{\rm exp}=0.0509 &\qquad& |V_{cb}|=0.034,\cr
BR(B^+\to D^{*0} l^+\nu)^{\rm exp}=0.0558&\qquad& |V_{cb}|=0.035.
\end{eqnarray}
 Thus the averaged $|V_{cb}|$  over all presented 
measurements of semileptonic decays $B\to De\nu$ and $B\to D^*e\nu$ is
equal to  
\begin{equation}
  \label{eq:bdavvcb}
  |V_{cb}|=0.040
\end{equation}
 and is in good agreement with  the experimental result \cite{pdg2012}.
\[|V_{cb}|=0.0396\pm0.0009\quad ({\rm exclusive}). \]

\section{Conclusion}
In this work, we have studied the renormalisation scale dependance of Isgur-wise function by using a wavefunction with linear part of the Cornell Potential as perturbation.  Considering the exclusive semileptonic decay of $B\to De\nu$ and $B\to D^*e\nu$, we then compute the CKM element$|V_{cb}|$ in this approach. The result of the CKM element is found to be within the error limits of other results. The following features are observed in this work:

\begin{itemize} 
\item  The renormalization scale of the model was set to be $\Lambda_{QCD}=410\;MeV$ as is used in ref.\cite{badalianalpha}.


\item The slope and curvature of the Isgur wise function are found to lie within the range of limits found in the literature. 
\item The computed value of the CKM element from the exlusive semileptonic decay of $B$ meson is obtained as $|V_{cb}|=0.040$. The result is to be well consistent with the lattice result $|V_{cb}|=0.0409\pm0.0015\pm0.0007$\cite{Divitis}.
\item  It becomes an worthy comment to note that the result with $\Lambda_{QCD}=200$(as is done in our previous work)is found to be very poor to study the branching ratio and $|V_{cb}|$.
  
\end{itemize}

From a phenomenological point of view, the present theoretical framework is a considerably useful tool to investigate various physical quantities for the heavy-light quark systems. Further work on the renormalisation schemes and scale dependance of strong coupling $\alpha_{s}$ is under study.


\begin{thebibliography}{99}
\bibitem{kkp5} K K Pathak, D K Choudhury and N S Bordoloi; Int.J.Mod.Phys.A vol.28(2013),1350010.\\
\bibitem{neubert1} M Neubert, Phys.Rep.245,259(1994)\\
\bibitem{kkp2} K K Pathak and D K Choudhury; Chinese Physics Lett.Vol.28,No.10(2011)101201\\
\bibitem{NSB2000} D K Choudhury and N S Bordoloi; IJMPA, vol.15; 23(2000)\\
\bibitem{Neubert}M Neubert ; IJMPA 11,4173 (1996)\\
\bibitem{Isgur}N Isgur and M B Wise;Phys.Lett.B 232,113(1989)\\
\bibitem{badalianalpha} A M Badalian and D S Kuzmenko, arXiv:hep-ph/0104097(2011)\\
\bibitem{Faustovalpha} D Ebert, R N Faustov and V O Galkin, Phys. Rev.D79:114029(2009)\\
\bibitem{Jugeau} F. JUGEAU etal;arXiv:hep-ph/0412144(2004)\\
\bibitem{younac}A. Le Yaouanc, L. Oliver and J.C. Raynal,Phys.Rev.D 69, 094022 (͑2004͒)\\
\bibitem{close} F E Close and A Wambach, Nucl.Phys.B412,169(1994)\\
\bibitem{NSB2009} D K Choudhury and N S Bordoloi; MPLA, vol. 24; 443(2009)\\
\bibitem{Faustovnewanalysis} D Ebert, R N Faustov and V O Galkin, Phys. Rev.D75:074008(2007)\\
\bibitem{Bowler}UKQCD Collaboration, K.C. Bowler et al., Nucl. Phys. B 637, 293 (2002).\\
\bibitem{ALEPH} D. Buskulic et al. (ALEPH collaboration), Phys. Lett. B395, 373–387 (1997)\\
\bibitem{Belle} K. Abe et al. (Belle collaboration), Phys. Lett. B526, 258–268 (2002),arXiv:hep-ex/0111082.\\
\bibitem{BDAI}Y B Dai ,C S Huang ,M K Huang  and C  Liu C; Phys.Lett.B 387;379 (1996)\\
\bibitem{IVANOV}M A Ivanov ,V E Lyubouvitskij ,L G K\"{o}rner ,P Kroll ; Phy.Rev.D 56;348(1997)\\
\bibitem{NEUBERT} M Neubert  ; Int.J.Mod.Phys.A 11,4173 (1996)\\
\bibitem{Qkim} Quang Ho-Kim and Phem Xuan Yem in"Elementary particles and their interaction"Springer publication,Germany.\\
\bibitem{pdg2012} J Beringer etal;(Particle Data Group),Phys.Rev.D86,010001(2012)\\
\bibitem{Divitis}G. M. de Divitiis, E. Molinaro, R. Petronzio, and N. Tantalo, Phys. Lett. B655, 45–49
(2007), arXiv:0707.0582 [hep-lat].





\end{thebibliography}
\end{document}